\documentclass[aps,prb,onecolumn,amsmath,superscriptaddress,amssymb,nofootinbib,preprint,11pt]{revtex4}
\usepackage{graphicx}
\usepackage{ulem} 
\usepackage{soul} 
\usepackage[usenames,dvipsnames]{color}  
\usepackage[colorlinks,citecolor=blue]{hyperref}
\usepackage{verbatim}
\usepackage{esint}
\usepackage{afterpage}
\renewcommand{\vec}[1]{\boldsymbol{#1}}
\def \k {{\vec k}}
\def \p {{\vec p}}

\def \r {{\vec r}}
\def \a {{\vec a}}

\def \q {{\vec q}}
\def \d{\partial}

\def \ll {\ell}

\def \K {{\vec K}}

\def \D{\Delta}
\def \A{{\bf A}}

\def \L{{\cal{L}}}
\def \O {\cal{O}}
\def \beq {\begin{eqnarray}}
\def \eeq {\end{eqnarray}}
\def \ben {\begin{equation}}
\def \een {\end{equation}}
\def \tn {\textnormal}

\begin{document}

\title{Higgs criticality in a two-dimensional metal}
\author{Debanjan Chowdhury}
\affiliation{Department of Physics, Harvard University, Cambridge Massachusetts-02138, USA.}
\author{Subir Sachdev}
\affiliation{Department of Physics, Harvard University, Cambridge Massachusetts-02138, USA.}
\affiliation{Perimeter Institute of Theoretical Physics, Waterloo Ontario-N2L 2Y5, Canada.}

\date{\today \\
\vspace{0.6in}}
\begin{abstract}
We analyze a candidate theory for the strange metal near optimal hole-doping in the cuprate superconductors.
The theory contains a quantum phase transition between metals with large and small Fermi surfaces of spinless fermions carrying the electromagnetic charge of the electron,
but the transition does not directly involve any broken global symmetries. The two metals have emergent SU(2) and U(1) gauge fields respectively,
and the transition is driven by the condensation of a real Higgs field, carrying a finite lattice momentum and an adjoint SU(2) gauge charge.
This Higgs field measures the local antiferromagnetic correlations in a `rotating reference frame'.
We propose a global phase diagram around this Higgs transition, and describe its relationship
to a variety of recent experiments on the cuprate superconductors.
\end{abstract}

\maketitle

\section{Introduction}
\label{sec:intro}

Several recent experiments \cite{JH14,JSD14,ZX14,Ramshaw14} have provided strong evidence for a dramatic change in the nature of the low temperature electronic
state of the hole-doped cuprate superconductors near optimal doping ($x=x_c$). Moreover, zero field photoemission experiments carried out in the normal state have seen evidence for a `large' Fermi-surface for $x>x_c$, consistent with the overall Luttinger count \cite{Damascelli05,Damascelli07}, and disconnected Fermi `arcs' near the nodal regions for $x<x_c$ \cite{Campuzano06}.  At high fields, quantum oscillations also reveal a `large' Fermi-surface for $x>x_c$ \cite{Proust08}, but a closed electron-like Fermi-surface with an area that constitutes a small fraction of the entire Brillouin-zone for $x<x_c$ \cite{LT07}. It is therefore quite natural to 
associate the transition with decreasing $x$ at $x=x_c$ with the loss of a `large' Fermi-surface and the simultaneous opening of a pseudogap. 
There has also been significant experimental 
progress \cite{MHJ11,Ghi12,DGH12,SH12,MHJ13,Greven14,comin2,SSJSD14,DGH14} in understanding
the structure of the density-wave ordering at lower doping, which is likely responsible for 
the reconstructed electron-like Fermi-surface seen in quantum oscillation experiments \cite{SSNHGL12, AADCSS14}. 

In this paper we will use these advances to 
motivate and develop a previously proposed model \cite{SS09} for the physics of the strange metal near optimal doping. We argue that the rich phenomenology observed in the underdoped cuprates is primarily driven by a transition between non-Fermi liquid metals with large and small Fermi surfaces which does not directly involve any broken global symmetry. 
All states with broken symmetry\footnote{We shall ignore the subtleties associated with the presence of quenched disorder, 
except when it acts as a source of momentum decay for DC transport, as discussed later.} observed at low temperatures and low doping
are not part of the critical field theory \cite{senthil1,balents},
but are derived as low energy instabilities of the parent small Fermi surface phase. 
This diminished role for broken symmetries is consistent with absence of any observed order with a significant correlation length at higher
temperatures.
We will also construct a global phase
diagram to describe the many phases and crossovers around the strange metal.

A quantum phase transition which does not involve broken symmetries is necessarily associated with a {\it topological} change in the character
of the ground state wavefunction. Emergent gauge fields are a powerful method of describing this topological structure, and they
remain applicable also to the gapless metallic phases of interest to us here. Given the fundamental connection between emergent gauge fields and the size of the Fermi surface, which was  established in 
Ref.~\onlinecite{SVS04} using Oshikawa's method \cite{oshikawa},
we are naturally led to a quantum phase transition in which
there is a change in the structure of the deconfined gauge excitations. Indeed, this describes 
a Higgs transition in a metal, such as that discussed in Ref.~\onlinecite{SS09}. This argument is a general motivation for Higgs criticality near optimal
doping in the cuprates, which applies beyond the specific model considered here.

We emphasize that we are using the traditional particle-physics terminology
in which a ``Higgs transition'' describes the breaking
of a local gauge invariance. We are not referring to the longitudinal mode of a broken global symmetry, which has also
been labeled ``Higgs'' in condensed matter contexts \cite{varma}. 

The primary new motivation for the model of Ref.~\onlinecite{SS09} arises from our recent work \cite{DCSS14} analyzing the $d$-form factor
density waves observed in scanning tunnelling microscopy \cite{SSJSD14} and X-ray experiments \cite{comin2}. 
In this work \cite{DCSS14}, we argued that such density waves 
arise most naturally as an instability of a metallic 
higher temperature pseudogap state with small Fermi surfaces 
described as a \cite{FFL,MPAASS15} `fractionalized Fermi liquid' (FL*); other works with related ideas on the 
pseudogap are Refs.~\onlinecite{Coleman89,Wen96,YRZ,DMFT09,LvvFL,MV12,Mei14}.
Specifically, we used a theory of the FL* involving 
a background U(1) spin liquid with bosonic spinons \cite{RK07,acl,YQSS10,PS12}: 
it is therefore convenient to dub this metallic state for the pseudogap
as\footnote{However the Fermi surface excitations in this FL* phase carry the same quantum numbers as the electron, and do not couple minimally to the emergent (deconfined) gauge-fields.} a U(1)-FL*. These results are also easily extended to a $\mathbb{Z}_2$ spin liquid, and we will consider this case in Appendix~\ref{app:spiral}. The presence of a small Fermi surface without symmetry breaking requires topological order and emergent gauge fields \cite{SVS04}, 
and so also a Higgs transition to the large Fermi surfaces at larger doping: here we provide a natural embedding of a FL* theory into such a transition,
and we expect similar approaches are possible for other possible topological orders in the underdoped regime.

\begin{figure}
\begin{center}
\includegraphics[width=4.5in]{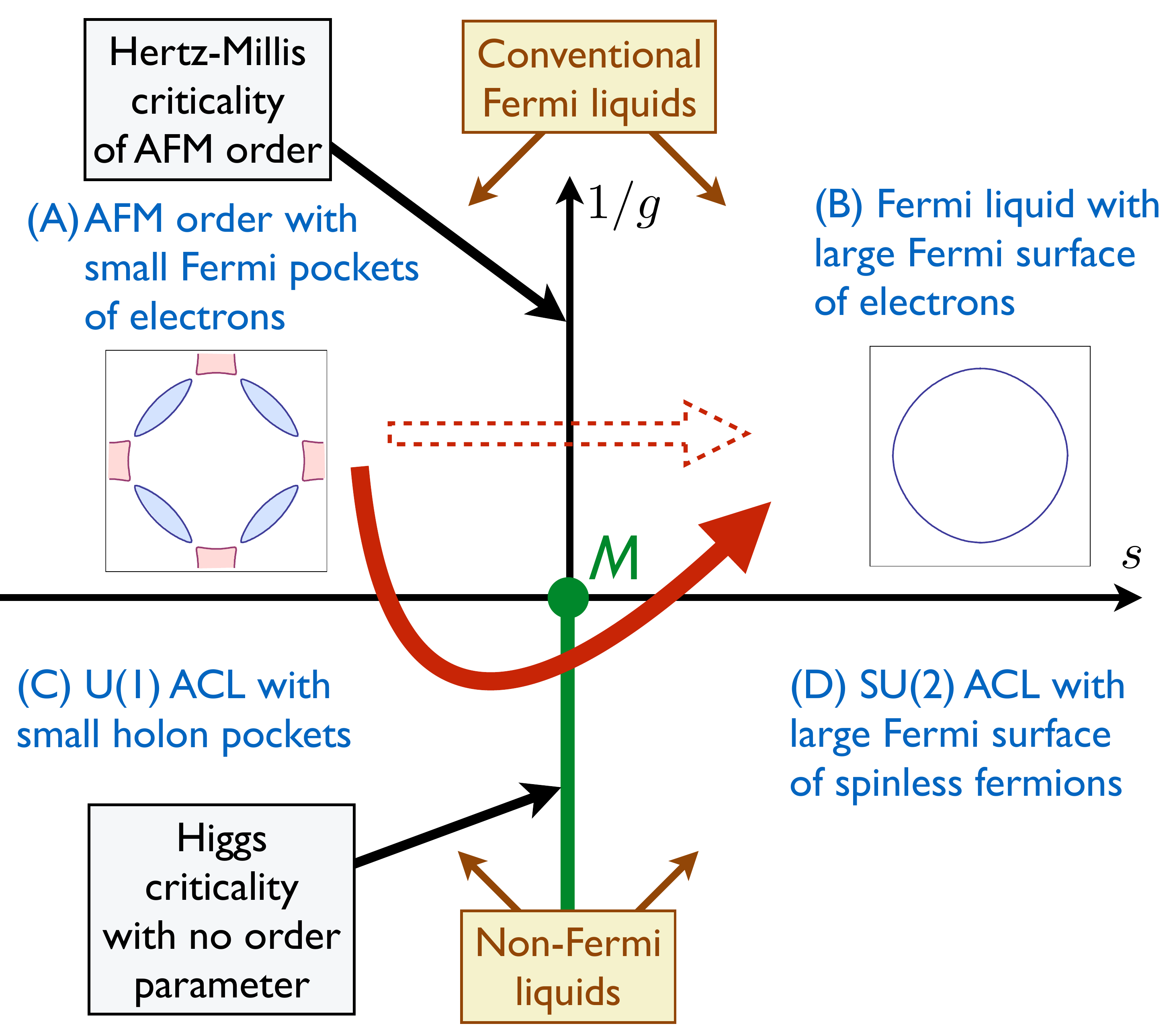}
\end{center}
\caption{Sketch of the metallic phases of our theory. Only phase A has a broken global symmetry, associated with the presence of long-range 
antiferromagnetic (AFM) order. The conventional Fermi liquid phases at the top have a transition from small to large Fermi surfaces
accompanied by the loss of AFM order. The dashed arrow represents a direct route between these phases, which could be a description of the electron-doped
cuprates. The full arrow around the point M is our proposed route with increasing doping in the hole-doped cuprates. The U(1)-FL* descends from the U(1) ACL,
as shown in Fig.~\ref{phasecup}. Note that the U(1)-FL* has a `small' Fermi surface of electrons due to the presence of topological order, while phase A above
has a `small' Fermi surface of electrons because of translational symmetry breaking.}
\label{fig:phase1}
\end{figure}

We now consider the evolution of the U(1)-FL*, and its small electronic Fermi surfaces, to the conventional `large' Fermi surface 
Fermi liquid state at large doping. There is an existing conventional theory of the transformation from
small to large Fermi surfaces driven by the disappearance of antiferromagnetic order. This is a transition between two Fermi liquids, and the
vicinity of the transition is described by the Hertz-Millis theory \cite{Hertz76,Millis93} 
and its field-theoretic extensions \cite{ChubukovShort,max1,strack,sungsik}, as shown in Fig.~\ref{fig:phase1}.
Here, we describe a detour from this direct route \cite{SS09} in which two new non-Fermi liquid phases appear between the conventional phases
of Hertz-Millis theory. The detour is described by a SU(2) gauge theory, and the transition from small to large Fermi surfaces is now a Higgs transition without any local order parameter, in which the emergent gauge structure
describing the topological order in the ground state changes from U(1) to SU(2). The Higgs field of this transition is a measure of the local antiferromagnetic
correlations in a rotating reference frame to be introduced below in Eq.~(\ref{R}).

Note that the Higgs transition in Fig.~\ref{fig:phase1} is between metallic states which we denote as `algebraic charge liquids' (ACL). 
The small and large Fermi surfaces in the ACLs are those of spinless fermions which carry the electromagnetic charge of the electron.
For the U(1) ACL, a bound state forms between the spinless fermions and a spin $S=1/2$ boson \cite{RK07,acl,YQSS10,PS12}, leading to 
small Fermi surfaces of fermionic quasiparticles carrying the same quantum numbers as the electron in the U(1)-FL*: so photoemission
will detect a small Fermi surface of electrons in the U(1)-FL*. We anticipate that similar effects are also present in the SU(2) ACL metal: there is a large
density of states of thermally excited $S=1/2$ bosons at low energy, 
so that the photoemission spectral function reflects the large Fermi surface of the spinless fermions.

We also note that although the Higgs field plays a central role in our phase diagram, its direct experimental detection will be difficult.
It is overdamped via its coupling to the Fermi surfaces, and gauge invariance prohibits any experimental probe from 
coupling linearly to it. Nevertheless, we will see below that it has significant experimental consequences via its strong effect on the fermionic spectrum.

\begin{figure}[ht]
\begin{center}
\includegraphics[scale = 0.65]{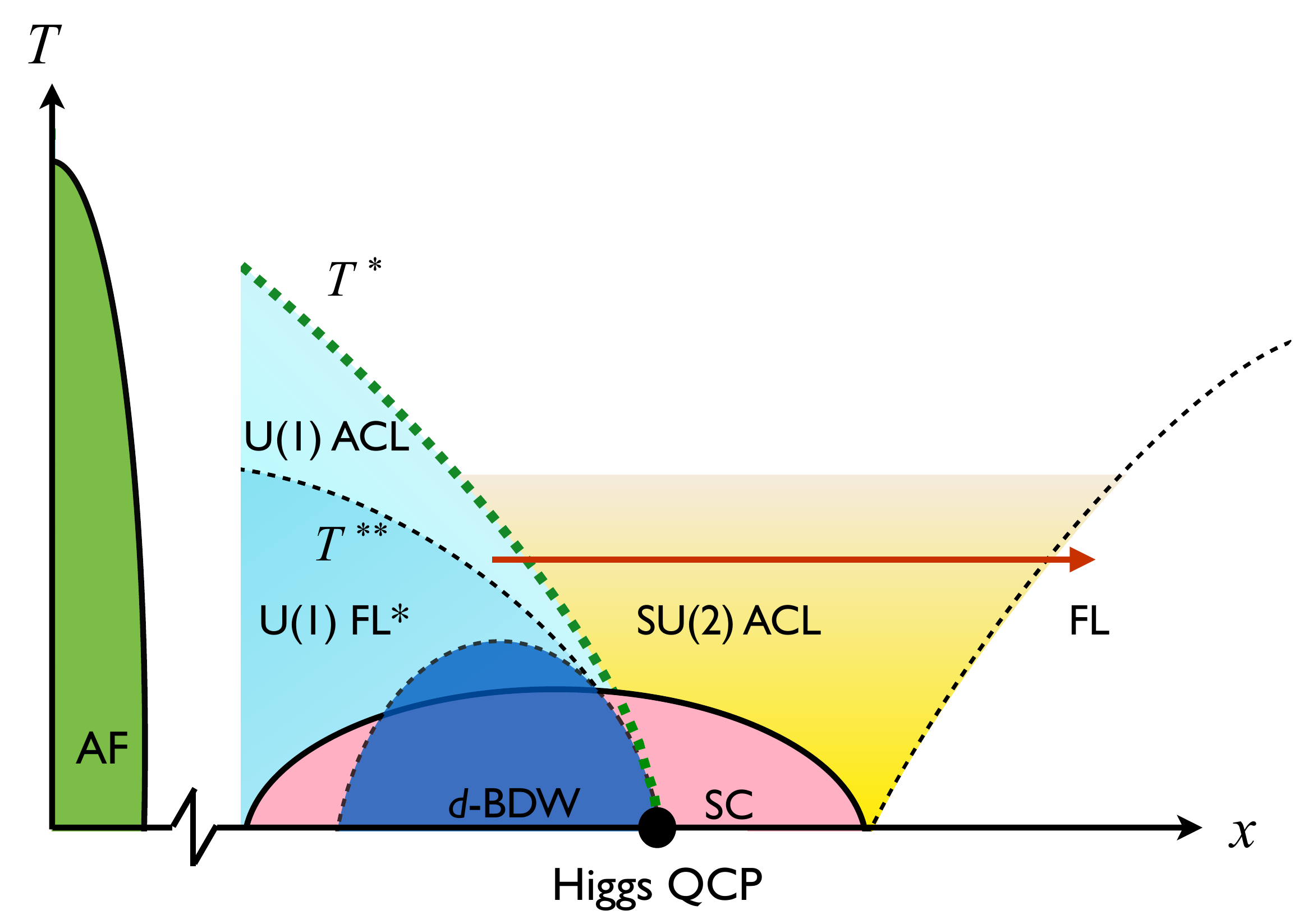}
\end{center}
\caption{Our proposed phase diagram for the hole-doped cuprates, building on a theory for Higgs criticality for the optimal
doping QCP. The green and red lines correspond to those in Fig.~\ref{fig:phase1}. The algebraic charge liquids (ACLs) have Fermi surfaces of spinless $\psi$ fermions which carry the electromagnetic charge:
in the SU(2) ACL the Fermi surface is `large' and is coupled to an emergent SU(2) gauge field, while in the U(1) ACL the
Fermi surface is `small' and coupled to an emergent U(1) gauge field.
The fractionalized Fermi liquid (FL*) descends from the U(1) ACL
by the binding of $\psi$ fermions to neutral spinons. The $d$-BDW is the $d$-form factor bond density wave, the SC is
the $d$-wave superconductor, and the FL is the large Fermi surface Fermi liquid. We are not concerned here with the physics of the extremely underdoped region. Also, we expect that the crossovers within the superconducting phase will 
exhibit a `back-bending' \cite{MS09,ZX12,ZX14} which is not shown above, and which we do not discuss further here.
The dashed lines at $T^\ast$ and $T^{\ast\ast}$ are crossovers, while the Higgs QCP at $T=0$ is a sharp phase transition.
}
\label{phasecup}
\end{figure}

We will present details of this theory starting from a microscopic model in Section~\ref{su2}, but first, in Section~\ref{sec:prelim}, 
we shall describe some key aspects using our proposed phase diagram in Fig.~\ref{phasecup}.

\section{Overview}
\label{sec:prelim}

Let us begin with a simplified picture of the optimal doping strange metal with a large Fermi surface. 
We consider a model of electrons $c_{i\alpha}$ on the
sites $i$ of a square lattice, with $\alpha = \uparrow, \downarrow$ a SU(2) spin index.
We transform the electrons to a rotating reference frame\footnote{This allows us to describe phases without long-range antiferromagnetic order.} using a SU(2) rotation $R_i$ and (spinless-)fermions 
$\psi_{i,p}$ with $p= \pm$,
\ben
\left( \begin{array}{c} c_{i\uparrow} \\ c_{i\downarrow} \end{array} \right) = R_i \left( \begin{array}{c} \psi_{i,+} \\ \psi_{i,-} \end{array} \right),
\label{R}
\een
where $R_i^\dagger R_i = R_i R_i^\dagger = 1$. Note that this representation immediately introduces a SU(2) gauge invariance (distinct from the global SU(2) spin rotation)
\ben
\left( \begin{array}{c} \psi_{i,+} \\ \psi_{i,-} \end{array} \right)\rightarrow U_i \left( \begin{array}{c} \psi_{i,+} \\ \psi_{i,-} \end{array} \right) \quad, \quad R_i \rightarrow R_i U_i^\dagger, \label{gauge}
\een
under which the original electronic operators remain invariant, $c_{i\alpha}\rightarrow c_{i\alpha}$; here $U_i$ is a  SU(2) gauge-transformation acting on the $p=\pm$ index. So the $\psi_p$ fermions are SU(2) gauge fundamentals, they carry the physical electromagnetic global U(1) charge, but they do not carry the SU(2) spin of the electron. The density of the $\psi_p$ is the same as that of the electrons.
Such a rotating reference frame perspective was used in the early work by Shraiman and Siggia on lightly-doped
antiferromagnets \cite{SS88,SS89}, and the importance of its gauge structure was clarified in Ref.~\onlinecite{SS09}.

The strange metal is obtained 
by forming a large Fermi surface state of the $\psi_p$ fermions, while $R_i$ fluctuate isotropically over all SU(2) rotations with a moderate
correlation length. This description suggests a simple trial wavefunction for this strange metal. Begin with a large Fermi surface (LFS) 
state of free $\psi_p$ fermions:
\ben
 \prod_{\k ~{\rm inside~ LFS},~ p=\pm} \psi_{p}^\dagger (\k)   \left| 0 \right\rangle .
\een
Expand this out in position space, insert the inverse of Eq.~(\ref{R}) to write the wavefunction in terms of $R$ and the physical electrons
$c_\alpha$, and finally average over $R$, to obtain
\ben
\int \prod_i dR_i ~W[ \{R_j\} ] \prod_{\k ~{\rm inside~ LFS},~ p=\pm} \left[ \sum_i e^{i \k \cdot \r_i} R_{i \alpha p} \, c_{i \alpha}^\dagger \right] 
\left| 0 \right\rangle,
\een
where $W$ is a variational weight-function of the $R_i$, invariant under global spin rotations.
For $W=1$, we have a zero correlation length for $R_i$, and we 
obtain a wavefunction for the $c_\alpha$ involving only empty and doubly-occupied sites. With non-trivial $W$, 
the correlation length of $R$ increases, we also build
in spin singlet pairs of $c_\alpha$ electrons on nearby sites. Comparing to the Gutzwiller-projected trial states commonly used for the
underdoped cuprates \cite{mohit}, this wavefunction includes the possibility of doubly-occupied sites and assigns different complex
weights to the off-site singlet pairs.

For a more precise and complete description of the strange metal, which accounts for the gauge structure in 
Eq.~(\ref{R}), we must turn to a quantum effective action for the $\psi_p$ which
necessarily includes an emergent SU(2) gauge field.
In the terminology of Ref.~\onlinecite{acl}, such a theory of spinless, gapless fermions coupled to an emergent gauge field is an
`algebraic charge liquid' (ACL), and hence we have labeled the strange metal as SU(2) ACL in Fig.~\ref{phasecup}.
This name implies that the SU(2) gauge symmetry is unbroken
({\it i.e.\/} not `Higgsed'), and in such a situation 
the $\psi_p$ fermions have a large Fermi surface with a shape similar to that of the electron Fermi surface in Fermi liquid state at large doping. 

Now let us consider the transition to the U(1) ACL in Fig.~\ref{phasecup}. This is described by the condensation of 
a real Higgs field $H^a$, where $a = 1,2,3$ indicates that the Higgs field transforms as a SU(2) adjoint. 
As we will see below in Eq.~(\ref{hn}), this Higgs field is a measure of the local antiferromagnetic order in the rotating reference
frame defined by $R$ (see also Ref.~\onlinecite{JM14} for an illuminating analogy).
The condensation of the Higgs field breaks the
gauge symmetry from SU(2) to U(1) and reconstructs the $\psi_p$ Fermi surface from large to small.
It is this Higgs transition which describes the optimal doping QCP in Fig.~\ref{phasecup}, and analyzing its structure is
the main purpose of the present paper. In the case where $H^a$ is complex, the Higgs phase can break the gauge symmetry down to $\mathbb{Z}_2$,
and we consider this case in Appendix~\ref{app:spiral}. The Shraiman-Siggia analyses \cite{SS88,SS89} of doped antiferromagnets were
effectively within such a Higgs-condensed regime, and this obscured the gauge structure of their formulation \cite{SS09}.

Let us also note from Fig.~\ref{phasecup} that the U(1) ACL is the parent of the U(1)-FL*.
This was discussed in Refs.~\onlinecite{acl,RK07}, and will be reviewed below: the U(1)-FL* arises by the formation of bound states between the spinons, $R$, and $\psi$ fermions around the small Fermi surface. We expect that a similar phenomenon also happens at low $T$ in the SU(2) ACL at lower temperatures, so that the photoemission largely reflects the structure of the
large Fermi surface of the SU(2) ACL.

The phase diagram in Fig.~\ref{phasecup} is meant to be schematic; determining the exact nature of the various crossover and phase-transition lines is beyond the scope of this work. The Higgs transition is present only at $T=0$, and there is only a crossover at $T>0$ (shown as the dashed green line in Fig.~\ref{phasecup}). Also, we have assumed that the energy scales associated with the ACL/FL* and $d-$BDW vanish with the same power law as a function of $(x-x_c)$ at the QCP\footnote{Corresponding to `case C' in Ref.~\onlinecite{TS14}.}. Theoretically speaking, other possibilities \cite{TS14} are also allowed.
\subsection{Field theory}
\label{sec:ft}

We now specify the imaginary time Lagrangian of the optimal doping QCP in Fig.~\ref{phasecup}, and its vicinity.
For now, the Lagrangian will not include the $R$ bosons: we assume that $R$ fluctuations are short-ranged, but the associated
spin-gap in the SU(2) ACL phase of Fig.~\ref{phasecup} is small because of proximity to the multi-critical point M in Fig.~\ref{fig:phase1};
we will include
the $R$ contributions in Section~\ref{su2}. Then we have,
\ben
\mathcal{L}_{\mathrm{QCP}} = \mathcal{L}_\psi + \mathcal{L}_H + \mathcal{L}_Y .
\label{LQCP}
\een

The first term describes a large Fermi surface of $\psi$ fermions minimally coupled to a 
SU(2) gauge field $A_\mu^a =(A_\tau^a ,\vec{A}^a)$:
\ben
\L_\psi = \sum_i \psi_{i,p}^\dagger[(\d_\tau-\mu)\delta_{pp'} + iA_\tau^a\sigma^a_{pp'} ]\psi_{i,p'} +  \sum_{i,j}t_{ij}\psi^\dagger_{i,p}\bigg[e^{i\sigma^a\vec{A}^a\cdot(\r_i-\r_j)}\bigg]_{pp'}\psi_{j,p'},
\een
where $t_{ij}$ are the fermion hopping parameters, $\r_i$ are the spatial co-ordinates of the sites, 
$\mu$ is the chemical potential, and $\sigma^a$ are Pauli matrices acting on
the SU(2) gauge indices. 

The Higgs Lagrangian is denoted $\mathcal{L}_H$, and it has a form familiar from its particle-physics
incarnations,
\ben
\L_H = \frac{1}{2} (\d_\tau H^a-2i\epsilon_{abc}A^b_\tau H^c)^2 + \frac{{\tilde{\rm{v}}}^2}{2} (\nabla H^a-2i\epsilon_{abc}\vec{A}^b H^c)^2+\frac{s}{2} (H^a)^2  + \frac{u}{24}[(H^a)^2]^2.
\label{LH}
\een
The Higgs potential is determined by the parameters $s$ and $u$, and transition across the QCP is controlled by
the variation in $s$. As usual, for negative $s$, the Higgs field condenses, and this breaks the gauge symmetry from
SU(2) to U(1); and for positive $s$, the Higgs field is gapped, and then the SU(2) gauge symmetry remains unbroken.

Finally, we have the Yukawa coupling in $\mathcal{L}_Y$. As in particle-physics, this is a trilinear coupling between
the Higgs field and the fermions, but now it has a different spatial structure:
\ben
\mathcal{L}_Y = -  \lambda \, H_i^a e^{i \K \cdot \r_i } \, \psi_{i,p}^\dagger \sigma^a_{pp'} \psi_{i,p'},
\label{yukawa}
\een
where $\K = (\pi,\pi)$ is the antiferromagnetic wavevector. This spatial structure indicates that $H^a$ transforms non-trivially
under lattice translations: 
\ben
H^a \rightarrow H^a e^{i \K \cdot \a}~\mbox{under translation by $\a$}; 
\label{Higgstrans}
\een
note that this is permitted
because $e^{i \K \cdot \a} = \pm 1$ is real for all spacings $\a$.
The transformation in Eq.~(\ref{Higgstrans}) arises from the role of the Higgs field as a measure of the antiferromagnetic correlations
in a rotating reference frame.
In the presence of the Higgs condensate, this Yukawa coupling
reconstructs the $\psi$ Fermi surface from large to small, and the $e^{i \K \cdot \r_i }$ factor is crucial in the structure of this reconstruction. 
While in the particle physics context the Higgs condensate
gives the fermions a mass gap, here the fermions acquire a gap only on certain portions of the large Fermi surface,
and a small Fermi surface of gapless fermions remains.

We note that the effective gauge theory will also acquire a Yang-Mills term for the SU(2) gauge field $A^a$ when high energy degrees of freedom
are integrated out. As is well known in theories of emergent gauge fields, such a term helps stabilize deconfined phases of the type considered here.
We do not write this term out explicitly here, but will include its contributions in Section~\ref{sec:landaudamping}, and specifically in the $L_A$ term
in Eq.~(\ref{eq:LSU2}).

\subsection{DC transport}
\label{trans}
The body of our paper will describe a field theoretic analysis of the non-Fermi liquid properties of 
$\mathcal{L}_{\mathrm{QCP}}$. This combines recent progress in the theories of Fermi surfaces coupled to order
parameters \cite{ChubukovShort,max1,strack,sungsik} and gauge fields \cite{SSLee,metnem,mross}. 
Here we mention one notable result on the electrical resistivity in the quantum-critical region
of the Higgs transition. As in recent work \cite{raghu,patel} on other quantum critical points of metals, we consider the situation in which there is a strong momentum bottleneck {\it i.e.\/} there is rapid exchange of momentum between the fermionic
and bosonic degrees of freedom, and the resistivity is determined by the rate of loss of momentum.
In particular, it is possible for the resistivity to be dominated by the scattering of neutral bosonic degrees of freedom,
rather than that of charged fermionic excitations near the Fermi surface.
In our model, we argue that an important source for momentum decay is the coupling of the Higgs field to disorder
\ben
\mathcal{L}_{\mathrm{dis}} =  V(\r) \left[ H^a (\r) \right]^2 , \label{Ldis}
\een
where $V(\r)$ is quenched Gaussian random variable with
\ben
\langle\langle V (\r )\rangle\rangle=0~~;~~\langle\langle V (\r) V (\r^{\prime})\rangle\rangle=V_0^2\delta^2(\r-\r^{\prime}),
\een
where the double angular brackets indicate an average over quenched disorder.
Comparing with Eq.~(\ref{LH}), we see that $V(\r)$ can be viewed as a random local variation in the value of $s$, the tuning
parameter which determines the position of the QCP. 
We will show that the analysis of the contribution of $\mathcal{L}_{\mathrm{dis}}$ to the resistivity closely parallels the computation in Ref.~\onlinecite{patel} for the spin-density-wave 
quantum critical point. And as in Ref.~\onlinecite{patel}, we find a resistivity for weak disorder which is proportional
to $V_0^2$, 
\beq
\rho(T)\sim V_0^2~ T^{2(\D+1-z)/z},
\label{dct}
\eeq
where $\D=d+z-\nu^{-1}$ is the scaling dimension of the $(H^{a})^2$ operator, $\nu$ is the correlation length exponent and $z$ is the dynamical exponent. 
As we will see in Section~\ref{sec:higgscritical}, this predicts a linear-in-$T$ resistivity for the leading order values of the exponents.

The outline for the rest of our paper is as follows. In Section \ref{su2}, we arrive at the above gauge-theoretic description starting from the theory of a metal with fluctuating antiferromagnetism and discuss the mean-field phase diagram as a function of the relevant tuning parameters. In Section \ref{qcp}, we describe the properties of the QCP using a low-energy description of the Fermi-surface coupled to a gauge-field and the critical fluctuations of the Higgs' field. Finally in Section \ref{conc}, we discuss the relation of our proposed phase-diagram to the actual phase-diagram in the hole-doped cuprates. Appendix~\ref{app:spiral} 
contains the extension to spiral order and $\mathbb{Z}_2$ gauge theory, while technical details are in Appendix~\ref{diags}.

\section{SU(2) gauge theory of antiferromagnetic metals}
\label{su2}

We summarize the derivation in Ref.~\onlinecite{SS09} of the SU(2) gauge theory, starting from a model of 
electrons on the square lattice coupled to the fluctuations of collinear antiferromagnetism at the wavevector $\K = (\pi, \pi)$. 
The case of collinear antiferromagnetism at other wavevectors was also considered in Ref.~\onlinecite{SS09}, and we treat
spiral antiferromagnets at incommensurate wavevectors in Appendix~\ref{app:spiral}. 

We begin with a model of electrons coupled to the quantum fluctuations of antiferromagnetism represented by
the unit vector $n_{i\ell}$, with $\ell = x, y, z$ and $\sum_\ell 
n_{i\ell}^2 = 1$. The Lagrangian is given by
\beq
\L&=&\L_f + \L_n + \L_{fn}, \nonumber \\
\L_f&=&\sum_i c_{i\alpha}^\dagger [(\d_\tau-\mu)\delta_{ij}-t_{ij}]c_{j\alpha}, \nonumber \\
\L_n&=&\frac{1}{2g} \bigg[(\d_\tau n_\ell )^2+{\rm{v}}^2(\nabla n_\ell)^2 \bigg], \nonumber \\
\L_{fn}&=&- \lambda\sum_i e^{i\K\cdot\r_i} ~n_{i\ell} \cdot c_{i\alpha}^\dagger \sigma^\ell_{\alpha\beta} c_{i\beta}.
\label{model}
\eeq
In the above $g$ measures the strength of quantum fluctuations associated with the orientation of $n_\ell$, $\lambda$ is an ${\O}(1)$ spin-fermion coupling and $\rm{v}$ is a characteristic spin-wave velocity.

Now we insert the parametrization in Eq.~(\ref{R}) into Eq.~(\ref{model}) and proceed to derive an effective theory for 
$\psi_{p}$ and $R$. The formulation of the latter theory is aided by the introduction of a SU(2) gauge 
connection $A_\mu^a =(A_\tau^a ,\vec{A}^a)$. As is familiar in many discussions of emergent gauge fields in 
correlated electron systems, this gauge field arises after decoupling hopping terms via an auxiliary field; here we skip
these intermediate steps, and simply write down appropriate hopping terms for the $\psi_{p}$ and $R$ 
which are made gauge-invariant by suitable insertions of the gauge connection. 

With the parameterization in Eq.~(\ref{R}) we notice that the coupling $\L_{fn}$ in Eq.~(\ref{model}) maps precisely onto
the Yukawa coupling in Eq.~(\ref{yukawa}) with
\ben
H_i^a \sigma^a_{pp'} = n_{i \ell} R^\ast_{i \alpha p} \sigma^\ell_{\alpha\beta} R_{i \beta p'},
\een
and so we {\it define\/} the Higgs field $H_{i}^a$ by
\ben
H_i^a \equiv \frac{1}{2} n_{i \ell} ~\tn{Tr}[\sigma^\ell R_i\sigma^a R_i^\dagger].
\label{hn}
\een
This identifies $H^a$ as the antiferromagnetic order in the rotating reference frame defined by Eq.~(\ref{R}).
An important property of this definition is that the field $H^a$ is {\it invariant} under a global SU(2) spin rotation $V$,
which rotates the direction of the physical electron spin and of the antiferromagnetic order,
\ben
\left( \begin{array}{c} c_{i\uparrow} \\ c_{i \downarrow} \end{array} \right)\rightarrow V \left( \begin{array}{c} c_{i\uparrow} \\ c_{i \downarrow} \end{array} \right) \quad, \quad R_i \rightarrow V R_i . \label{spin}
\een
Note that the SU(2) spin rotation is a left multiplication of $R$ above, while the SU(2) gauge transformation in Eq.~(\ref{gauge}) is a right
multiplication of $R$. With these properties, Eq.~(\ref{hn}) implies that $H^a$ transforms as a vector under the SU(2) gauge
transformation in Eq.~(\ref{gauge}).

We have now assembled all the steps taken after substituting Eq.~(\ref{R}) into Eq.~(\ref{model}).
The Lagrangian of the resulting gauge theory is then obtained as
\ben
\mathcal{L}_{\mathrm{SU(2)}} = \L_{\mathrm{QCP}} + \L_R,
\label{su2l}
\een
where $\L_{\mathrm{QCP}}$ was described below Eq.~(\ref{LQCP}) in Section~\ref{sec:ft},
and $\L_R$ is the Lagrangian for $R$. The structure of the latter is determined by the transformations of $R$ in Eqs.~(\ref{gauge})
and Eq.~(\ref{spin}). So we have
\ben
\L_R =  \frac{1}{2g}\tn{Tr}\bigg[(\d_\tau R - iA_\tau^aR\sigma^a)(\d_\tau R^\dagger + iA_\tau^a\sigma^aR^\dagger) + {\rm{v}}^2(\nabla R - i\vec{A}^aR\sigma^a)(\nabla R^\dagger + i\vec{A}^a\sigma^aR^\dagger) \bigg].
\een
This completes our derivation of the SU(2) gauge theory.

It is useful here to collect the transformations of the fields under the SU(2) gauge transformation, the global SU(2) spin rotation,
and electromagnetic U(1) charge, as summarized in table \ref{tab}.
\begin{table}
\begin{tabular}{c|c|c|c}
& $\mathrm{SU(2)}_{\mathrm{gauge}}$ & $\mathrm{SU(2)}_{\mathrm{spin}}$ & $ \mathrm{U(1)}_{\mathrm{e.m. charge}}$ \\
\hline
$c$ & {\bf 1} & {\bf 2} & 1 \\
$n$ & {\bf 1} & {\bf 3} & 0 \\
$\psi$ & {\bf 2} & {\bf 1} & 1 \\
$H$ & {\bf 3} & {\bf 1} & 0 \\
$R$ & ${\bf \overline{2}}$ & {\bf 2} & 0
\end{tabular}
\caption{Summary of the transformations of the fields under the various gauge-transformations. SU(2) representations of spin $S$ 
are labelled by their dimension of $2S+1$. The U(1) column contains the charge under the U(1) gauge field.}
\label{tab}
\end{table}

Finally, we can make contact with other approaches by expressing $R$ as
\beq
R_i =\left( \begin{array}{cc} 
z_{i\uparrow} & -z_{i\downarrow}^*\\
z_{i\downarrow} & z_{i\uparrow}^*
\end{array}
\right), \label{Rz}
\eeq
with $|z_{i \uparrow}|^2 + |z_{i \downarrow}|^2 = 1$, but this parameterization will not be useful to us.
Consider the situation in the Higgs phase, where the field $H^a$ is condensed. Then we are free to choose a gauge
in which the Higgs condensate is $H^a = (0,0,1)$. In such a condensate, after inverting the relation in Eq.~(\ref{hn})
we find
\beq
n_{i \ell} &=& \frac{1}{2} H_i^a ~\tn{Tr}[\sigma^\ell R_i\sigma^a R_i^\dagger] \nonumber \\
&=& z_{i \alpha}^\ast \sigma^\ell_{\alpha\beta} z_{i \beta} \quad \mbox{for $H^a = (0,0,1)$}.
\label{e1}
\eeq
The last relationship is the familiar connection between the O(3) and CP$^1$ variables, but note
that it holds here only within the phase where the Higgs field is condensed {\it i.e.\/} in the U(1) ACL.

\subsection{Mean field phase diagram}

\begin{figure}[ht!]
\begin{center}
\includegraphics[scale = 0.4]{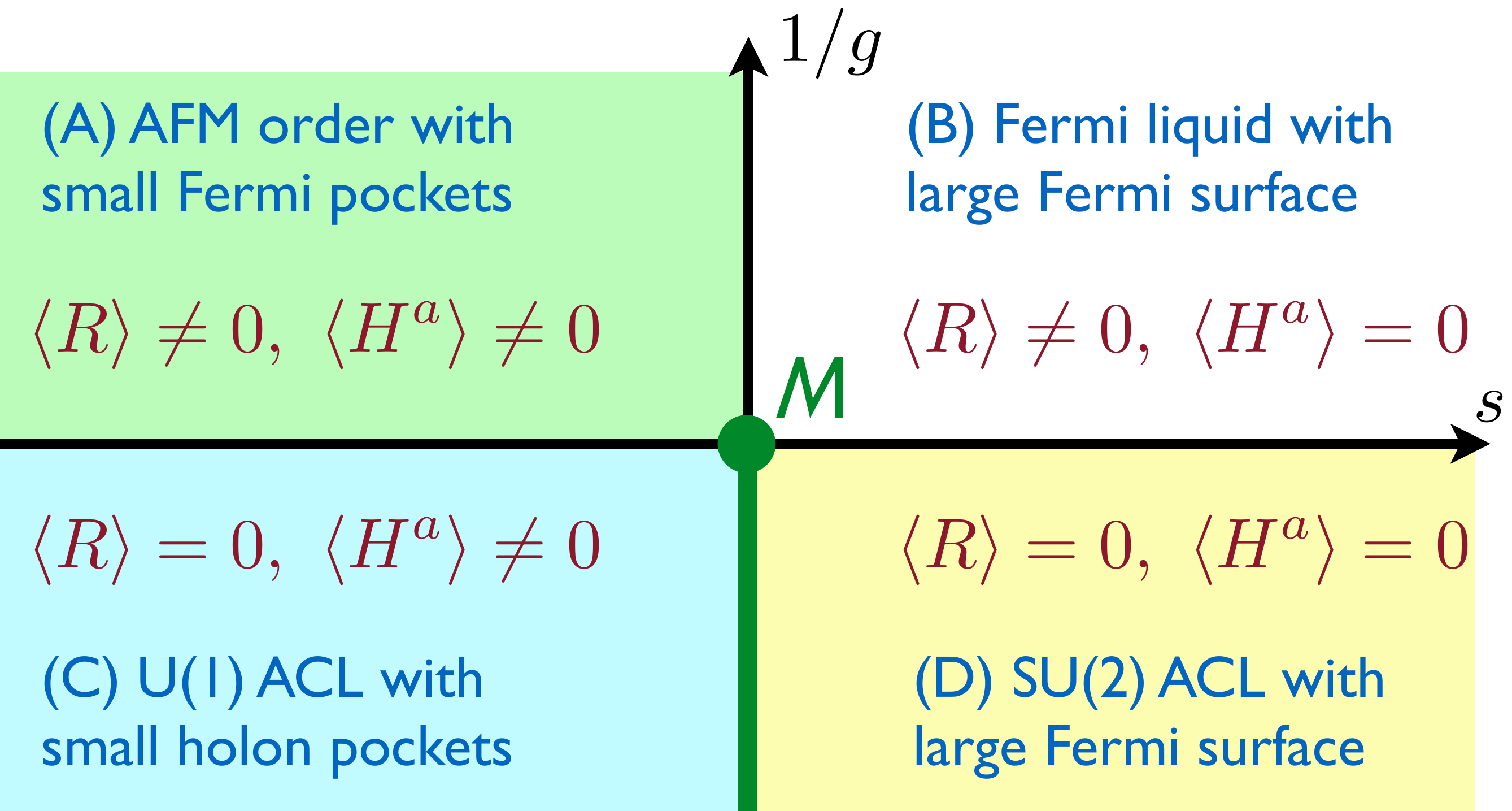}
\end{center}
\caption{The phase diagram for the theory in Eq.~(\ref{su2l}) as a function of $s$ and $1/g$ (also shown in Fig.~\ref{fig:phase1}). 
The color-coding of the phases corresponds to that in Fig.~\ref{phasecup}. 
The multicritical point, M, corresponds to $g=g_c$ and $s=0$. This paper is concerned with the critical properties associated with the transition (C)$\leftrightarrow$(D).}
\label{fig:phase2}
\end{figure}

We now describe the phases of $\L_{\mathrm{SU(2)}}$ obtained in a simple mean field theory \cite{SS09}
in which we allow condensates of the bosonic field $R$ and $H^a$. 
These phases are obtained by varying the tuning parameters $s$ and $g$, and were shown in 
Fig.~\ref{fig:phase1}; in Fig.~\ref{fig:phase2}, we label the phases by their condensates.
The phases are:
\begin{itemize}
\item The Higgs phase, labelled as (A) in Figs.~\ref{fig:phase1},\ref{fig:phase2}, where both SU(2)$_\tn{spin}$ and SU(2)$_\tn{gauge}$ are broken, leading to $\langle R\rangle\neq0,~\langle H^a\rangle\neq0$. The gauge-excitations, $(A_\tau,\vec{A})$, are gapped here. This phase describes the {\it AFM-metal} where the large Fermi-surface gets reconstructed into hole (and electron) pockets due to condensation of $H^a\sim\vec{n}$, the N$\acute{\tn{e}}$el order parameter. 

\item The SU(2) confining phase, labelled as (B) in Figs.~\ref{fig:phase1},\ref{fig:phase2}. Note that the SU(2)$_\tn{spin}$ here remains unbroken. We have $\langle R\rangle\neq0,~\langle H^a\rangle=0$, which is necessary to preserve spin-rotation invariance since $\vec{n}=0$ from Eq.~(\ref{hn}). This is the usual {\it Fermi liquid} phase, with a large Fermi-surface.

\item The Higgs phase, labelled as (C) in Figs.~\ref{fig:phase1},\ref{fig:phase2}, where the SU(2)$_\tn{gauge}$ is broken, but the SU(2)$_\tn{spin}$ remains unbroken, leading to $\langle R\rangle=0,~\langle H^a\rangle\neq0$. By recalling the physical interpretation of the fields, this amounts to a locally well developed amplitude of the AFM, without any long-range orientational order. We can choose $H^a\sim(0,0,1)$ by carrying out a gauge-transformation, which immediately implies  that a U(1) subgroup of the SU(2)$_\tn{gauge}$ remains unbroken, so that the $A^z$ photon remains gapless. Thus this phase describes a {\it U(1) algebraic charge liquid}, or, the {\it holon-metal} \cite{acl}. However, due to the locally well developed AFM order, the Fermi-surface is reconstructed into $\psi_p$  holon pockets that are minimally coupled to a U(1) gauge-field. 

As a function of temperature, there could be a continuous crossover from a U(1) ACL to a U(1) FL$^*$ (or a ``holon-hole" metal), where some of the holons ($\psi_{\pm}$) start forming bound states with the gapped spinons ($z_{\alpha}$) \cite{acl}.

\item The final phase (D) in Figs.~\ref{fig:phase1},\ref{fig:phase2} has the full symmetry, with none of the fields condensed: $\langle R\rangle = \langle H^a\rangle=0$. Instead of the above U(1) ACL, where only $A^z$ was gapless, in this phase there are a triplet of gapless SU(2) photons coupled to a large Fermi-surface. This phase can be described as a {\it SU(2) algebraic charge liquid}. Formally, this phase a spin gap, but we assume that $T$ is greater
than the gap in the metallic regions of Fig.~\ref{phasecup} because of proximity to the point M in Fig.~\ref{fig:phase1}. At low enough $T$, this phase is
unstable to superconductivity \cite{MMTS}.
\end{itemize}
We should emphasize that the above mean-field analysis has been rudimentary; {\it e.g.\/} we cannot rule out the possibility
that higher order couplings could induce first-order transitions, that could even eliminate an intermediate phase.

The next section shall present the theory for the interplay between the fluctuations of the gauge and Higgs' fields, within a low-energy field-theoretic formulation. 

\section{Low-energy field theory} 
\label{qcp}

We are interested in studying the properties of the QCP between the SU(2) ACL and the U(1) ACL. At the QCP, $s=0$, 
the entire Fermi-surface is coupled to the transverse fluctuations of a SU(2) gauge field. There have been studies in the particle physics
literature of Fermi surfaces coupled to non-Abelian gauge fields \cite{schafer1,schafer2}; 
however these have been restricted to spatial dimension $d=3$, where
a RPA analysis gives almost the complete answer. In spatial dimension $d=2$ of interest to us here, we shall follow the approach taken
for Abelian gauge theories \cite{SSLee,metnem,mross} which uses a patch decomposition of the Fermi surface. The same approach
transfers easily to the non-Abelian case; indeed because of the Landau damping of the gauge bosons, there is little difference
between the Abelian and non-Abelian cases \cite{SS09,schafer1,schafer2}, as will also be clear from our analysis in Section~\ref{sec:landaudamping}.

Apart from their coupling to a SU(2) gauge field, the fermionic $\psi_p$ particles are also coupled to a quantum critical Higgs field.
This coupling is strongest at 8 `hot spots' around the Fermi surface, and in Section~\ref{sec:higgscritical} 
we shall be able to use the methods developed
from the case of a spin-density-wave transition of Fermi liquids \cite{ChubukovShort,max1,strack,sungsik}

Some of the details of the computations appear in Appendix \ref{diags}.

\begin{figure}[ht!]
\begin{center}
\includegraphics[scale = 0.45]{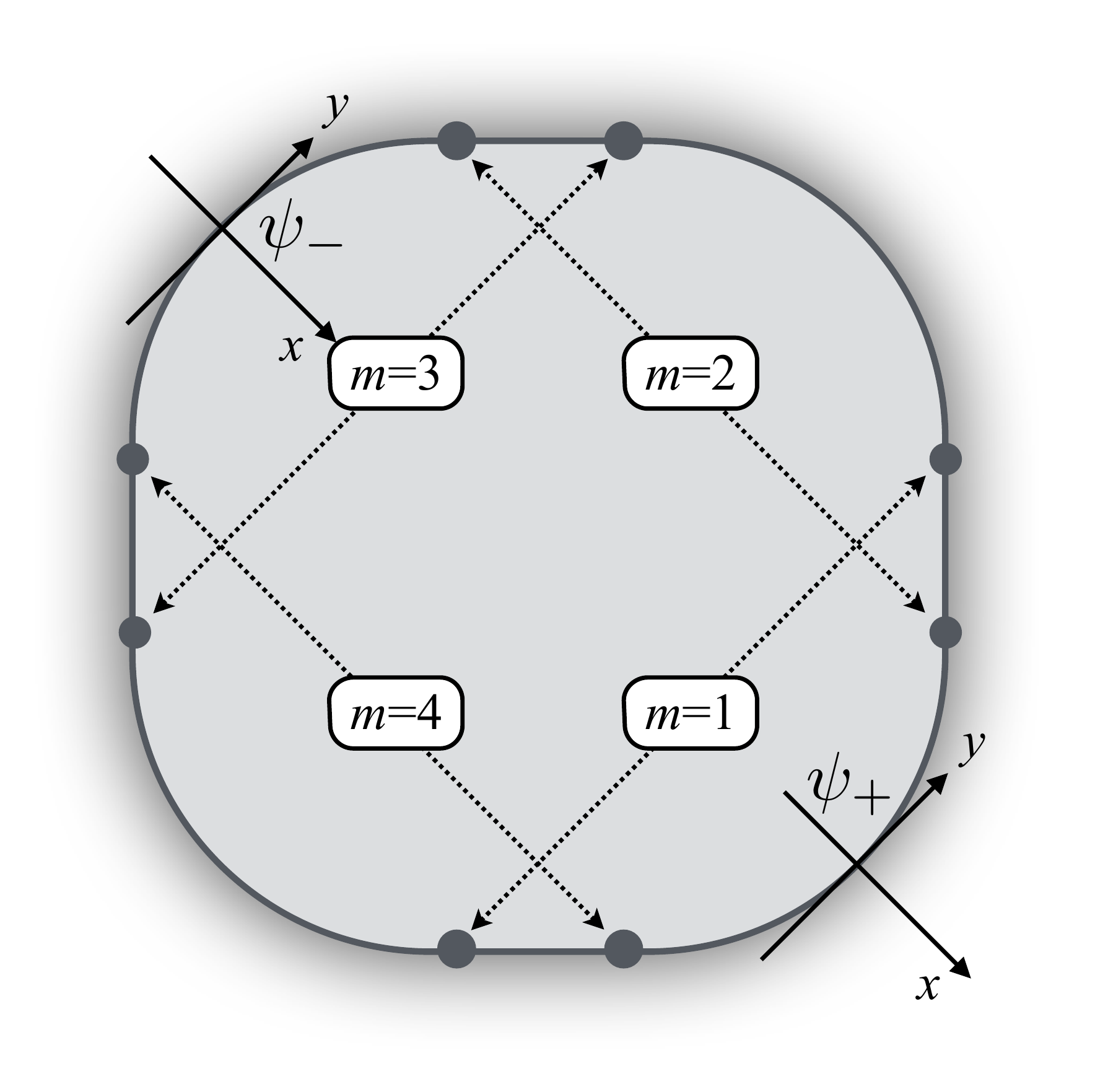}
\end{center}
\caption{The shaded grey regions represent the occupied states. The transverse gauge-field fluctuations couple strongly to the flavor current arising from the $\psi_\pm$ patches and destroy the $\psi$ quasiparticles all around the Fermi-surface. Across the Higgs transition, the fluctuations of the $H^a$ field couple most strongly to the four-pairs ($m=1,..,4$) of ``hot-spots", shown as the filled circles.}
\label{FS}
\end{figure}

\subsection{Fermi-surface coupled to gauge-field} 
\label{sec:landaudamping}

Here we describe the low energy theory of the SU(2) ACL, away from the Higgs condensation at the QCP.
We need only consider a SU(2) gauge field coupled to the large Fermi surface of the $\psi_p$ fermions. As in the U(1) case \cite{SSLee,metnem,mross}, we
can make a patch decomposition of the Fermi surface, and treat antipodal pairs of patches separately. 
For a single pair of antipodal patches, we have the fermions $\psi_{\pm p}$ (see Fig.~\ref{FS}), with $\pm$ the patch index, and $p$ the usual
SU(2) gauge index. This is coupled to the transverse components of the SU(2) gauge field, $\A^a$.

\beq
L&=&L_f+L_\A+L_{\tn{int}}, \nonumber \\
L_f&=&\psi_{+p}^\dagger(\partial_\tau-i\partial_x-\partial_y^2)\psi_{+p} + \psi_{-p}^\dagger(\partial_\tau+i\partial_x-\partial_y^2)\psi_{-p}, \nonumber \\
L_\A&=&\frac{1}{2e^2} (\partial_y A^a_x)^2,\nonumber \\
L_\tn{int}&=& A^a_x (\psi_{+p}^\dagger \, \sigma^a_{pp'}  \, \psi_{+p'} - \psi^\dagger_{-p} \, \sigma^a_{pp'} \, \psi_{-p'}) \label{eq:LSU2}
\eeq

\begin{figure}[ht!]
\begin{center}
\includegraphics[scale = 0.8]{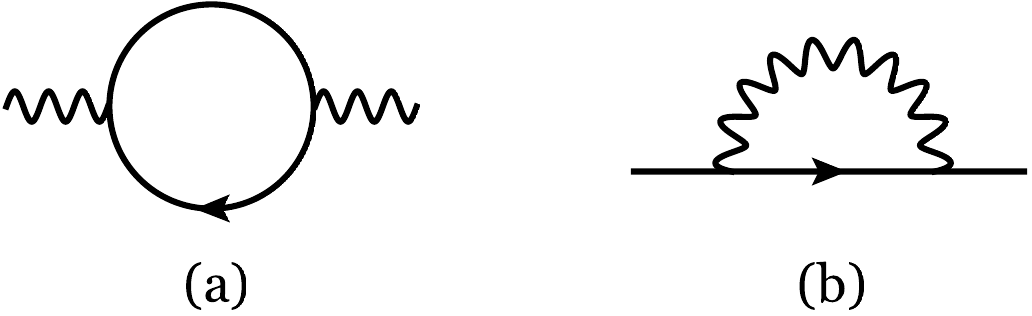}
\end{center}
\caption{One loop contributions to the (a) gauge-field, and, (b) Fermion self-energies. Curly lines represent the $\A$ propagators, $D(\ll)$, while solid lines represent the $\psi$ propagators, $G(\ll)$.}
\label{gSE}
\end{figure}

Let us review the one-loop renormalization of the gauge and fermionic matter fields. We start by looking at the self-energy of the gauge-field due to the particle-hole bubble (Fig.~\ref{gSE}a). We have,
\beq
\Pi^\A_0(q)=2 \sum_s\int\frac{d\ll_\tau d^2\vec \ll}{(2\pi)^3} ~G_s^0(\ll)~G_s^0(\ll+q),
\label{gfs}
\eeq
where $\ll=(\ll_\tau,\vec{\ll})$ and the bare fermionic propagator is given by,
\beq
G_s^0(\ll)=\frac{1}{-i \ll_\tau + s\ll_x + \ll_y^2}.
\eeq
The final result is of the form\footnote{We note that since the fermions are strictly in two-dimensions, the non-universal factor of $\Lambda$, the UV cutoff, drops out. The factor that appears in general is of the form $\Lambda^{d-2}$, where $d$ is the number of space-dimensions.},
\beq
\Pi^\A_0(q)=c_b\frac{|q_\tau|}{|q_y|},~~\tn{where}~c_b=\frac{1}{2\pi}.
\label{pigf}
\eeq
The computations are summarized in Appendix \ref{gsea}.

Computing the fermionic self-energy due to the bosonic-propagator dressed with the RPA level polarization bubble (Fig.~\ref{gSE}b) leads to,
\beq
\Sigma_{s,pp'}(k)&=&-\sigma_{p\alpha}^a \sigma_{\alpha p'}^a \int \frac{d\ll_\tau d^2\vec{\ll}}{(2\pi)^3} ~D(\ll)~ G_s^0(k-\ll),\\
&=&-3~\delta_{pp'} \int \frac{d\ll_\tau d^2\vec{\ll}}{(2\pi)^3} ~D(\ll)~ G_s^0(k-\ll),
\eeq 
where $D(\ll)$ is the gauge-field propagator,
\beq
D^{-1}(\ll)=\bigg(c_b\frac{|\ll_\tau|}{|\ll_y|} + \frac{1}{e^2}\ll_y^2  \bigg).
\eeq
We then obtain,
\beq
\Sigma_s(k)&=&-\frac{3i}{2}\int\frac{d\ll_\tau d\ll_y}{(2\pi)^2}\frac{\tn{sgn}(k_\tau-\ll_\tau)}{c_b|\ell_\tau|/|\ll_y|+\ll_y^2/e^2},\\
\Sigma_s(k)&=&-ic_f~\tn{sgn}(k_\tau)|k_\tau|^{2/3},~~\tn{where}~c_f=2\sqrt{3}\bigg(\frac{e^2}{4\pi} \bigg)^{2/3}.
\label{sig1}
\eeq
This self-energy contribution is larger than the bare $\partial_\tau$ term at low energies. Therefore, upon including the RPA contribution into the fermionic propagator, we have,
\beq
G_s(\ll)=\frac{1}{-ic_f\tn{sgn}(\ll_\tau)~|\ll_\tau|^{2/3} + s\ll_x +\ll_y^2},
\label{fg}
\eeq
which is the well known result for the quasiparticles being damped all along the Fermi-surface.

\subsection{Higgs criticality at the QCP}
\label{sec:higgscritical}
 
Now we consider the QCP at which the Higgs boson condensed from the non-Fermi liquid SU(2) ACL state described in the previous
subsection. Across this Higgs transition from the SU(2) ACL to the U(1) ACL, the Fermi-surface gets reconstructed---this is controlled by the real Higgs field, $H^a$, which carries lattice momentum, $\K=(\pi,\pi)$. 
By the same arguments used for the onset of spin-density-wave order in a Fermi liquid \cite{ChubukovShort,max1,strack,sungsik}, the
low energy physics of the QCP is dominated by the vicinity of the hot-spots: these are points on the Fermi surface which are connected
by $\K$ (see Fig.~\ref{FS}).  The computation for the present non-Fermi liquid Fermi surface proceeds just as for the Fermi liquid case,
by linearizing the bare dispersion for the fermions around the hot spots:
\beq
L&=&L_{hs}+L_H+L_{fH},\nonumber \\
L_{hs}&=&\psi_{1p}^{\dagger m}(\partial_\tau - i\vec{v}^m_1\cdot\nabla)\psi_{1p}^m + \psi_{2p}^{\dagger m}(\partial_\tau - i\vec{v}^m_2\cdot\nabla)\psi_{2p}^m,\nonumber \\
L_{fH}&=&\frac{1}{\sqrt{N_f}}H^a\cdot(\psi_{1p}^{\dagger m}\sigma^a_{pp'}\psi_{2p'}^m + \psi_{2p}^{\dagger m}\sigma^a_{pp'}\psi_{1p'}^m),
\eeq 
where $L_H$ already appeared in Eq.~(\ref{LH}); $m$ is the hot-spot pair index (Fig.~\ref{FS}). 

\begin{figure}[ht!]
\begin{center}
\includegraphics[scale = 0.8]{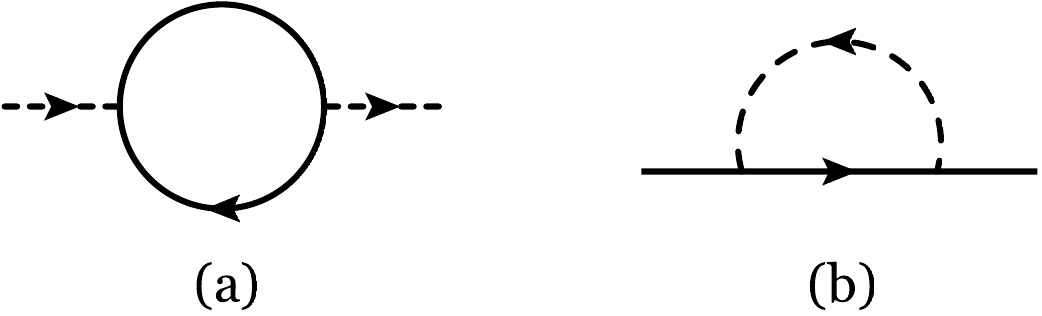}
\end{center}
\caption{One loop contributions to the (a) Higgs-field, and, (b) Fermion self-energies. The dashed lines represent the Higgs' field propagator, $\chi$.}
\label{hSE}
\end{figure}

Let us first look at the one-loop self energy of the $H^a$ field (Fig.~\ref{hSE}a). This is given by,
\beq
\Pi^H(q)=2\sum_m\int\frac{d\ll_\tau d^2\vec\ll}{(2\pi)^3} \bigg[G_1^m(\ll+q) G_2^m(\ll) + G_2^m(\ll+q) G_1^m(\ll) \bigg],
\label{pih}
\eeq
where we now use the non-Fermi liquid 
fermion Green's function renormalized by the gauge field fluctuations, as discussed in Section~\ref{sec:landaudamping}:
\beq
G_\alpha^m(\ll)=\frac{1}{-ic_f\tn{sgn}(\ll_\tau) |\ll_\tau|^{2/3} -\vec{v}_\alpha\cdot\vec\ll}.
\label{gf1}
\eeq
Note the $z=3/2$ scaling of the fermion self energy, which allows us to drop the  bare frequency dependence from above ($\sim\partial_\tau$). \\

Upon including contributions from all pairs of hot-spots, we obtain (see Appendix \ref{apph}),
\beq
\Pi^H(q)=\Pi^H(q=0)+\gamma |q_\tau|, ~~\tn{where}~\gamma=\frac{n}{2\pi v_xv_y},
\label{pih1}
\eeq
where $n=4$ is the number of pairs of hot spots.
Note that the $c_f$ dependence has completely dropped out and the above result is precisely the expression that we would have obtained if we had started with the bare fermion Green's functions (or, any anomalous power $\sim |\ll_\tau|^\beta$). This result is not surprising---it just reproduces the ``Landau-damped" form of the propagator for $H^a$. As we know, the only requirement for the appearance of Landau-damping is the existence of particle-hole excitations around the Fermi-surface in the limit of $\omega\rightarrow0$. In the general case, this always leads to $\sim |q_\tau|/|q_y|$ for a bosonic order-parameter coupled to a fermion-bilinear. When the order parameter itself carries a finite momentum $\K$, as is the case here, then the denominator in the damping term gets cut off and leads to $\sim |q_\tau|$.

Equipped with the above expression, let us now compute the self-energy of the fermions in the vicinity of the hot-spots (Fig.~\ref{hSE}b),
\beq
\Sigma_{1,pp'}(p)&=&\sigma_{p\alpha}^a\sigma_{\alpha p'}^a \int \frac{d\ll_\tau d^2\vec\ll}{(2\pi)^3} G_2(p-\ll) \chi(\ll),\\
&=&3\delta_{pp'}\int \frac{d\ll_\tau d^2\vec\ll}{(2\pi)^3} G_2(p-\ll) \chi(\ll),
\label{fhse}
\eeq
where the propagator tuned to the critical point ($s=0$) is given by : $\chi^{-1}(\ll)=(\gamma|\ll_\tau| + \vec\ll^2)$. 

We are interested in the singular power law frequency dependence of the self-energy at the hot-spots. 
For future use, it is useful to express the Green's function in 
Eq.~(\ref{gf1}) in the more general form
\beq
G_\alpha^m(\ll)=\frac{1}{-i\zeta_f\tn{sgn}(\ll_\tau) |\ll_\tau|^{\beta} -\vec{v}_\alpha\cdot\vec\ll},
\label{gf2}
\eeq
where the exponent $\beta=2/3$ from the coupling to the SU(2) gauge field.

Upon evaluating the momentum integrals, the self-energy (for $\p=0$) becomes (see Appendix \ref{fse}),
\beq
\Sigma_1(p_\tau)=3i\int \frac{d\ll_\tau}{4\pi^2}~ \tan^{-1}\bigg(\frac{\sqrt{\gamma v_2^2|\ll_\tau|-\zeta_f^2|p_\tau-\ll_\tau|^{2\beta}}}{\zeta_f|p_\tau-\ll_\tau|^{\beta}}\bigg)\frac{\tn{sgn}(\ll_\tau-p_\tau)}{\sqrt{v_2^2\gamma |\ll_\tau|-\zeta_f^2|p_\tau-\ll_\tau|^{2\beta}}},
\label{sef}
\eeq
which correctly reproduces $\Sigma_1(p_\tau=0)=0$. Furthermore, note that if $\zeta_f=0$, i.e. if the anomalous self-energy contribution were to be absent, then the above reduces to the well known form \cite{max1}
\beq
\Sigma_1(p_\tau) = 3i\int \frac{d\ll_\tau}{8\pi}\frac{\tn{sgn}(\ll_\tau-p_\tau)}{\sqrt{v_2^2\gamma |\ll_\tau|}} = -\frac{3i}{2\pi\sqrt{\gamma v_2^2}}~ \tn{sgn}(p_\tau) \sqrt{|p_\tau|},
\label{z2}
\eeq
reproducing the $z=2$ result. Let us now proceed to evaluate the expression in the presence of a finite $\zeta_f$. 
Rescaling $\ll_\tau=x p_\tau$ leads to,
\beq
&& \Sigma_1(p_\tau) 
=\frac{3i}{2\pi\sqrt{\gamma v_2^2}}~ \tn{sgn}(p_\tau) |p_\tau|^{1-\beta} \nonumber \\
&& ~~~\times \int \frac{dx}{2\pi}~ \tan^{-1}\bigg(\frac{\sqrt{|p_\tau|^{1-2\beta} |x|-c|1-x|^{2\beta}}}{\sqrt{c}|1-x|^\beta}\bigg)\frac{\tn{sgn}(x-1)}{\sqrt{|p_\tau|^{1-2\beta} |x|-c|1-x|^{2\beta}}},
\label{scself}
\eeq
where the dimensionless parameter, $c=\zeta_f^2/\gamma v_2^2$. An asymptotic analysis of Eq.~(\ref{scself}) shows that
\beq
\Sigma_1 (|p_\tau| \rightarrow 0) \sim -i \, \tn{sgn}(p_\tau) ~\sqrt{|p_\tau|} \quad , \quad \mbox{for $\beta \geq 1/2$.}
\eeq
So the low energy singularity of the self-energy is {\it independent\/} of the non-Fermi liquid exponent $\beta$ in the fermion Green's function,
and has the same value as in the spin density wave case without the gauge field. This is the key observation of the present 
subsection. To estimate the co-efficient, we can use a self-consistent approach in which we use a self energy in Eq.~(\ref{gf2}) with $\beta = 1/2$.
This self-energy arises from the coupling to the Higgs field, and is always dominant over the one obtained from the gauge field
with $\beta = 2/3$. Assembling all the constraints, the final expression takes the following $z=2$ form
\beq
\Sigma_1(p_\tau)=\frac{3i}{2\pi\sqrt{\gamma v_2^2}}~{\cal{I}}(c)~\tn{sgn}(p_\tau) \sqrt{|p_\tau|},
\eeq 
where the function $\mathcal{I}(c)$ is defined in Appendix~\ref{ints} and ${\cal{I}}(c\rightarrow0)=-1$.
 
So we reach our main conclusion that, in both the fermionic and bosonic sectors, the low energy physics of the Higgs QCP
is essentially identical to that of the spin-density-wave onset transition in a Fermi liquid. And the basic reason for this is simple.
The hotspot theory has dynamic critical exponent $z=2$, while the singularities arising from the SU(2) gauge field coupling
around the Fermi surface have $z=3/2$. At a given length scale, the contributions with the larger $z$ dominate
because they have a lower energy. Hence the Higgs criticality of a non-Fermi liquid maps onto the spin density wave criticality
of a Fermi liquid. 

With this conclusion in hand, we can now directly apply the results of Ref.~\onlinecite{patel} on the DC resistivity to the Higgs
QCP. The approach of Ref.~\onlinecite{patel} requires that there is quasiparticle breakdown around the entire Fermi surface,
and the fermionic excitations rapidly equilibriate with all the bosonic modes. While this was only marginally true for the spin-density-wave
quantum critical point considered in Ref.~\onlinecite{patel}, it is easily satisfied for the Higgs QCP being considered here:
the SU(2) gauge field makes the entire Fermi surface ``hot'', while the Higgs field fluctuations induce additional fermion damping at the hot spots
on the Fermi surface. 
As in the previous case, it is possible for disorder to couple to the square of the Higgs field because such an operator
is gauge-invariant, as we noted in Eq.~(\ref{Ldis}). And the corresponding contribution to the resistivity is in Eq.~(\ref{dct}).
For the exponents $d=2$, $z=2$, and $\nu=1/2$ presented above, this yields a linear-in-temperature results $\rho (T) \sim V_0^2 T$.

\section{Discussion}
\label{conc}

The primary goal of this paper has been to propose a candidate theory for the quantum phase transition near optimal doping in the cuprates.   We analyzed the QCP between metals with `large' and `small' Fermi-surfaces, which did not involve any broken global symmetries, but instead involved a Higgs' transition between metals with emergent SU(2) and U(1) gauge fields. 
The Higgs field acts as a measure of the local antiferromagnetic order in the rotating reference frame defined by Eq.~(\ref{R}).
As we discussed in Sections~\ref{sec:intro} and~\ref{sec:prelim}, the symmetry broken phases observed in the underdoped cuprates arise as low temperature instabilities of the `small' Fermi-surface metal. 

The underlying QCP we studied was between two metals (the U(1) ACL and the SU(2) ACL) 
in which the Fermi surface excitations are coupled to emergent gauge fields,
and so there are no Landau quasiparticles. However, electron-like quasiparticles do re-emerge around a small Fermi in the U(1)-FL*, and we will discuss
similar features around the large Fermi surface in the SU(2) ACL below.
The reconstruction to the `small' Fermi-surface in the ACL phases is driven by the condensation of the Higgs field, and the Higgs critical point has  additional singular structure in the vicinity of the ``hot-spots". The Higgs criticality has associated with it an interplay of both $z=3/2$ physics on the whole Fermi-surface, and $z=2$ physics in the vicinity of the hot-spots. We showed that near the Higgs QCP the $z=2$ physics
dominates, and hence many critical properties map onto the previously studied problem of the onset of spin density wave
order in a Fermi liquid \cite{ChubukovShort,max1,strack,sungsik,patel}.

Let us now conclude with a discussion of the relationship of our proposed phase diagram in Fig.~\ref{phasecup} to the experimentally obtained phase diagram in the non-La-based cuprates. The $d-$SC and $d-$BDW both arise as instabilities of the U(1) FL*, as has been discussed in 
Refs.~\onlinecite{DCSS14, EGSS11} (the SU(2) ACL is also unstable to superconductivity \cite{MMTS}). 
The high temperature pseudogap phase for $T^{**}<T<T^*$ is a U(1) ACL or more appropriately a {\it holon-hole} metal \cite{acl}. There is a crossover to the U(1) FL* phase at $T=T^{**}$, where all the holons have formed bound-states with the spinons. 
An important feature of the U(1) FL* phase is that its transport and photoemission signatures are mostly identical to those of a Fermi liquid.
The primary difference from a Landau Fermi liquid is that the volume enclosed by the Fermi surface is proportional to the density of holes, $x$, 
and not to the Luttinger density $1+x$. The U(1) FL* phase also has an emergent U(1) gauge field, as required by the topological arguments in Ref.~\onlinecite{SVS04}, but the Fermi surface quasiparticles are gauge neutral. The recent remarkable observation of Fermi liquid transport
properties in the pseudogap phase of Hg1201 \cite{Marel13, MG14} below $T^{**}$, with some possibly non-Fermi liquid behavior between $T^{**}$ and $T^{*}$, can therefore be viewed as strong support for the existence of a FL* derived out of a parent ACL.
In particular, the alternative `fluctuating order' picture of the pseudogap does not naturally lead to such temperature dependent crossovers from non-Fermi liquid to Fermi liquid regimes.

For the La-based cuprates, there is a larger doping regime with magnetic order, overlapping with the regime of charge order. This can be accommodated
in our phase diagram \cite{DCSS14} by moving the full red arrow in Fig.~\ref{fig:phase1} just to the other side of the point M, and allowing for incommensurate order as in Appendix~\ref{app:spiral}.

An important challenge for future experiments is to detect direct 
experimental signatures of the complete small Fermi surface of the proposed FL* phase. We presume that it is the small quasiparticle residue on the `back side' of the small Fermi surface \cite{YQSS10,MPAASS15} which is responsible for the arc-like features in the photoemission spectrum \cite{Yang11}. Therefore, we need a probe which does not involve adding or removing an electron from the sample, and so is not sensitive
to the quasiparticle residue. Possibilities are Friedel oscillations, the Kohn anomaly, or ultrasonic attenuation. 

Within our proposed phase diagram in Fig.~\ref{phasecup}, the strange metal phase is to be viewed as a SU(2) ACL at the Higgs critical point,
and proximate to the multicritical point M to ensure the spin gap is smaller than $T$.
The DC transport properties of this phase are controlled by the coupling of the gauge-invariant square of the Higgs field to {\it long\/} wavelength disorder, following an analysis of Ref.~\onlinecite{patel} for the spin density wave critical point.
Such a coupling leads to a linear-in-temperature resistivity.
Also, as emphasized in Ref.~\onlinecite{patel}, the residual resistivity is proportional to {\it short \/} 
wavelength disorder which can scatter fermions across the Fermi surface. 
So there is no direct relationship between the residual resistivity and the slope of the linear resistivity.
It would be interesting in future work to explore the role of intrinsic
umklapp scattering events in the transport properties of such strange metals in the strong coupling regime.

The electron spectral function in the SU(2) ACL is a convolution of the spectra of the $\psi$ fermions and the $R$ bosons.
As in the computation in Ref.~\onlinecite{RK07}, we assume the $R$ spectrum is thermally overdamped (because of the proximity to M), 
and the electron
spectral function primarily reflects the $\psi$ spectrum; we also expect precursors of the bound state formation between
the $\psi$ and the $R$ to enhance the $\psi$ features in the electron spectrum \cite{RK07}, just as in the U(1)-FL*. Then the electron spectral functions should have an
anisotropic structure around the Fermi surface, with 
the weaker gauge field-induced damping in the nodal region, and the stronger Higgs field-induced damping in the anti-nodal region.
Also note that while the Higgs field coupling does show up in the resistivity as discussed above, the gauge fields coupling 
has a weaker effect on transport. This is because gauge-invariance prevents a non-derivative coupling between the gauge field
and perturbations that violate momentum conversation. An important open question is whether
this rich theoretical structure can be made consistent with the complex experimental features of the conductivity and magnetotransport 
in the strange metal \cite{NH08,HKMS,Blake15}.

Our linear-$T$ resistivity is proportional to disorder, as in the previous model in Ref.~\onlinecite{patel}. However, because the disorder couples
to the Higgs field, the relevant disorder is long-wavelength. This is in contrast to short wavelength disorder, which can
lead to efficient large momentum scattering of fermions around the Fermi surface. Modifying the coefficient of the resistivity therefore requires
modifying long-wavelength disorder, and this may be difficult to do because of the intrinsic disorder from the dopant ions. 
Inducing short-wavelength disorder, by including {\it e.g.\/} Zn impurities, may not be effective in modifying the co-efficient
of the linear-$T$ resistivity. These features can act as tests of our proposed mechanism for the resistivity of the strange metal \cite{Alloul}.

Finally, we note from Figs.~\ref{phasecup} and~\ref{fig:phase1},\ref{fig:phase2} that the SU(2) ACL survives for an extended region
beyond the Higgs QCP. This implies strange metal behavior over a finite range of doping as $T \rightarrow 0$, and not only
at a single QCP. Transport measurements \cite{NH08} in magnetic fields which have suppressed superconductivity 
appear to be consistent with such a non-Fermi liquid phase.

\acknowledgements
We thank Seamus Davis and Martin Greven for useful discussions. 
D.C. is supported by the Harvard-GSAS Merit fellowship.
This research was supported by the NSF under Grant DMR-1360789, the Templeton foundation, and MURI grant W911NF-14-1-0003 from ARO.
Research at Perimeter Institute is supported by the Government of Canada through Industry Canada 
and by the Province of Ontario through the Ministry of Research and Innovation. 

\appendix

\section{Spiral order and $\mathbb{Z}_2$ gauge theory}
\label{app:spiral}

Here we generalize the theory in Eq.~(\ref{model}) to the case of spiral spin order at an incommensurate wavevector $\K$.
In this case the antiferromagnetic order is not characterized by a single 
unit vector $n_\ell$, but by two orthogonal unit vectors $n_{1\ell}$ and $n_{2\ell}$ which obey
\ben
n_{1\ell}^2 = n_{2 \ell}^2 = 1 \quad, \quad n_{1 \ell} \, n_{2 \ell} = 0. \label{triad}
\een
The spin-fermion coupling to the electrons in Eq.~(\ref{model}) is replaced by
\ben
\L_{fn} = - \lambda\sum_i \left[ n_{i 1\ell} \cos \left( \K\cdot\r_i \right) +  
n_{i 2\ell} \sin \left( \K\cdot\r_i \right)  \right]
\cdot c_{i\alpha}^\dagger \sigma^\ell_{\alpha\beta} c_{i\beta},
\een
After the change of variables in Eq.~(\ref{R}), this leads to the Yukawa coupling
\ben
\L_Y = -  \lambda \, \frac{1}{2} \left( H_i^{a \ast}  e^{i \K \cdot \r_i } + H_i^{a}  e^{-i \K \cdot \r_i } \right) \, \psi_{i,p}^\dagger \sigma^a_{pp'} \psi_{i,p'}, \label{yukawac}
\een
where, in contrast to Eq.~(\ref{yukawa}), the Higgs field $H^a$ is now {\it complex} and is defined by
\ben
H^a \equiv \frac{1}{2} \left(n_{1\ell} + i n_{2 \ell} \right) ~\tn{Tr}[\sigma^\ell R_i\sigma^a R_i^\dagger],
\label{hnc}
\een
generalizing Eq.~(\ref{hn}).
It is now also clear from Eq.~(\ref{yukawac}) that under translation by a distance $\a$, the Higgs field transforms as
in Eq.~(\ref{Higgstrans}), where $e^{i \K \cdot \a}$ can now be complex.

The structure of SU(2) gauge theory with a complex Higgs field remains essentially the same as for the real Higgs discussed
in the body of the paper, with one important distinction. The quartic term in Eq.~(\ref{LH}) is replaced by two terms
\ben
u_1 [H^{a \ast} H^a]^2 + u_2 [H^{a}]^2 [H^{b\ast}]^2 , \label{u12}
\een
and the presence of spiral order requires that $u_2 > 0$. Then in the Higgs phase, the minimum energy 
condensate can always be oriented so that
\ben 
H^a = (1,i, 0).
\een
Such a Higgs condensate breaks the SU(2) gauge symmetry all the way down to $\mathbb{Z}_2$. And using Eq.~(\ref{Rz}), 
the analog of the relationship
in Eq.~(\ref{e1}) for the orientation of the spiral order is
\beq
n_{1\ell} + i n_{2 \ell} &=& \frac{1}{2} H^a ~\tn{Tr}[\sigma^\ell R\sigma^a R^\dagger] \nonumber \\
&=& - \varepsilon_{\alpha\gamma} z_\gamma \sigma^\ell_{\alpha\beta} z_{\beta} \quad \mbox{for $H^a = (1,i,0)$}.
\label{e2}
\eeq
This co-incides with the conventional representation \cite{qpt} of the spiral orientation in terms of spinons $z_\alpha$, and it
can verified that the values in Eq.~(\ref{e2}) obey Eq.~(\ref{triad}).

For the case with $u_2 < 0$ in Eq.~(\ref{u12}), the Higgs condensate is instead a SU(2) rotation of
\ben 
H^a = e^{i \theta} (1,0, 0),
\een
where $\theta$ is an arbitrary phase. This corresponds to incommensurate collinear spin order \cite{SS09}.

\section{Feynman diagram computations}
\label{diags}
\subsection{Self-energy: Gauge-field }
\label{gsea}
Since we are interested in the singular structure of $\Pi^\A_0(q)$ in Eq.~(\ref{gfs}), we 
shall evaluate the integral over $\ll_x$ first, followed by $\ll_\tau, \ll_y$. Therefore,
\beq
\Pi^\A_0(q)&=& 2 \int \frac{d\ll_\tau d\ll_y}{(2\pi)^2}~\frac{i[\theta(\ll_\tau) - \theta(\ll_\tau+q_\tau)]}{-i\eta q_\tau + q_x + q_y^2 + 2\ll_yq_y } + \vec{q}\rightarrow-\vec{q},\\
&=&\frac{q_\tau}{\pi}\int \frac{d\ll_y}{(2\pi)}\frac{-i}{-i\eta q_\tau+q_x+q_y^2+2\ll_yq_y} + \vec{q}\rightarrow-\vec{q},\\
&=&\frac{|q_\tau|}{2\pi |q_y|}.
\eeq
This leads to the expression for $\Pi^\A_0(q)$ in Eq.~(\ref{pigf}).

\subsection{Self-energy: Higgs' field}
\label{apph}
Focusing on just the $m=1$ contribution, Eq.~(\ref{pih}) becomes,
\beq
\Pi^H_{m=1}(q)&=&2\int\frac{d\ll_\tau d^2\vec\ll}{(2\pi)^3} \bigg[ \frac{1}{-ic_f\tn{sgn}(\ll_\tau+q_\tau) |\ll_\tau+q_\tau|^{2/3}-\vec{v}_1\cdot(\vec\ll+\vec{q})}~\frac{1}{-ic_f\tn{sgn}(\ll_\tau) |\ll_\tau|^{2/3}-\vec{v}_2\cdot\vec\ll} \nonumber\\
&&~~~~~~~~~~~~~~~~~+ q\rightarrow-q \bigg].
\eeq 
Let us define $\ll_1= \vec{v}_1\cdot(\vec\ll + \q)$ and $\ll_2=\vec{v}_2\cdot\vec\ll$, so that
\beq
\Pi^H_{m=1}(q)=\frac{1}{v_xv_y}\int\frac{d\ll_\tau d\ll_1 d\ll_2}{(2\pi)^3} \bigg[ \frac{1}{-ic_f\tn{sgn}(\ll_\tau+q_\tau) |\ll_\tau+q_\tau|^{2/3}-\ll_1}~\frac{1}{-ic_f\tn{sgn}(\ll_\tau) |\ll_\tau|^{2/3}-\ll_2} \nonumber\\
+ q\rightarrow-q \bigg].
\eeq

It is not hard to see that the only non-zero contribution comes from the imaginary parts of both the terms. Then,
\beq
\Pi^H_{m=1}(q)=\frac{1}{v_xv_y}\int\frac{d\ll_\tau d\ll_1 d\ll_2}{(2\pi)^3} \bigg[ \frac{ic_f\tn{sgn}(\ll_\tau+q_\tau) |\ll_\tau+q_\tau|^{2/3}}{c_f^2 |\ll_\tau+q_\tau|^{4/3} + \ll_1^2}~\frac{ic_f\tn{sgn}(\ll_\tau) |\ll_\tau|^{2/3}}{c_f^2|\ll_\tau|^{4/3}+\ll_2^2} + q\rightarrow-q \bigg].
\eeq
Upon carrying out the $\ll_1,~\ll_2-$ integrals, this becomes,
\beq
\Pi^H_{m=1}(q)=-\frac{1}{4v_xv_y}\int\frac{d\ll_\tau}{2\pi}[\tn{sgn}(\ll_\tau+q_\tau)~\tn{sgn}(\ll_\tau) + q\rightarrow-q].
\eeq
This directly leads to the expression for $\Pi^H(q)$ in Eq.~(\ref{pih1}).

\subsection{Fermion self-energy at the hot-spot}
\label{fse}
The Fermionic self-energy due to the Higgs' field fluctuations (Eq.~(\ref{fhse})) becomes,
\beq
\Sigma_1(p)=3 \int \frac{d\ll_\tau d^2\vec\ll}{(2\pi)^3} \frac{1}{-i\zeta_f\tn{sgn}(p_\tau-\ll_\tau) |p_\tau-\ll_\tau|^{\beta} - \vec{v}_2\cdot(\p-\vec\ll)} ~\frac{1}{\gamma|\ll_\tau|+\vec\ll^2}.
\eeq
Let us now change coordinates such that $\ll_\perp=\hat{v}_2\cdot\vec\ll$ and $\ll_\Vert$ is the component along the Fermi-surface of $\psi_2$. Then,
\beq
\Sigma_1(p)=-3 \int \frac{d\ll_\tau d\ll_\perp d\ll_\Vert}{(2\pi)^3} \frac{1}{-i\zeta_f\tn{sgn}(p_\tau-\ll_\tau) |p_\tau-\ll_\tau|^{\beta} - \vec{v}_2\cdot\p + v_2\ll_\perp} ~\frac{1}{\gamma|\ll_\tau|+\ll_\perp^2+\ll_\Vert^2}.
\eeq
It is straightforward to carry out the integral over $\ll_\Vert$, which gives,
\beq
\Sigma_1(p)=-\frac{3}{2} \int \frac{d\ll_\tau d\ll_\perp}{(2\pi)^2} \frac{1}{-i\zeta_f\tn{sgn}(p_\tau-\ll_\tau) |p_\tau-\ll_\tau|^{\beta} - \vec{v}_2\cdot\p + v_2\ll_\perp} ~\frac{1}{\sqrt{\gamma|\ll_\tau|+\ll_\perp^2}}.
\eeq
Let us now study the form of the self-energy at the hot-spot, $\p=0$, and extract the $p_\tau$ dependence. We can symmetrize the above form then to give,
\beq
\Sigma_1(p_\tau)=3i\int \frac{d\ll_\tau}{2\pi} \int_0^\infty \frac{ d\ll_\perp}{2\pi}\frac{\zeta_f\tn{sgn}(\ll_\tau-p_\tau) |p_\tau-\ll_\tau|^{\beta}}{\zeta_f^2 |p_\tau-\ll_\tau|^{2\beta} + v_2^2\ll_\perp^2} ~\frac{1}{\sqrt{\gamma|\ll_\tau|+\ll_\perp^2}}.
\eeq
Carrying out the integral over $\ll_\perp$ (see Appendix \ref{ints}) leads to the expression in Eq.~(\ref{sef}).
\subsection{Integrals}
\label{ints}
\begin{figure}[h!]
\begin{center}
\includegraphics[scale = 0.25]{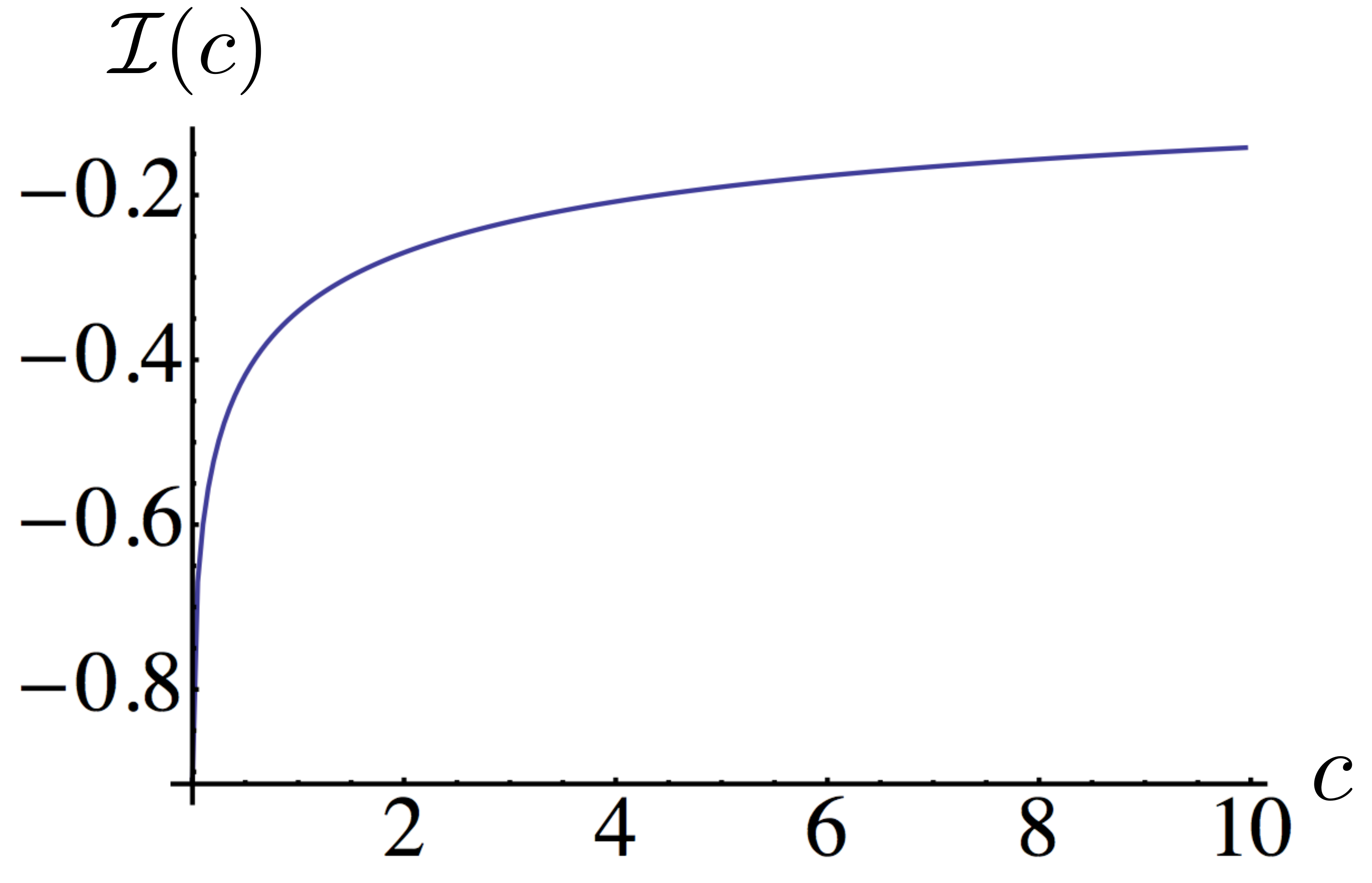}
\end{center}
\caption{${\cal{I}}(c)$ evaluated numerically as a function of $c$. Note that ${\cal{I}}(c\rightarrow0)=-1$, which reproduces Eq.\ref{z2}.}
\label{ic}
\end{figure}
We use the integral (for $a>0, b>0$):
\beq
\int_0^\infty dx\frac{1}{x^2+a^2}\frac{1}{\sqrt{x^2+b^2}} = \frac{1}{a\sqrt{b^2-a^2}}\tan^{-1}\bigg(\frac{\sqrt{b^2-a^2}}{a} \bigg).
\eeq
The above is valid irrespective of whether $a>b$ or $a<b$. 

The integral in Eq.~(\ref{scself}) can be evaluated as a function of the dimensionless parameter, $c=\zeta_f^2/\gamma v_2^2$, when $\beta=1/2$. The integral becomes,
\beq
{\cal{I}}(c)=\int_{-\infty}^\infty \frac{dx}{2\pi}~ \tan^{-1}\bigg(\frac{\sqrt{ |x|-c|1-x|}}{\sqrt{c|1-x|}}\bigg)\frac{\tn{sgn}(x-1)}{\sqrt{ |x|-c|1-x|}}.
\eeq

We show the functional form of ${\cal{I}}(c)$ as a function of $c$ in Fig.~\ref{ic}.

\end{document}